\documentclass[doublecol]{epl2}

\title{Entangled states of trapped ions allow measuring the magnetic field gradient  of a single atomic spin}
\author{F. Schmidt-Kaler\inst{1}, R. Gerritsma\inst{1}}

\institute{
  \inst{1} QUANTUM, Institut f\"{u}r Physik, Universit\"{a}t Mainz, D-55128 Mainz, Germany}

\abstract{Using trapped ions in an entangled state we propose detecting a magnetic dipole of a single atom at distance of a few $\mu$m. This requires a measurement of the magnetic field gradient at a level of about 10$^{-13}$~Tesla/$\mu$m. We discuss applications e.g. in determining a wide variation of ionic magnetic moments, for investigating the magnetic substructure of ions with a level structure not accessible for optical cooling and detection,and for studying exotic or rare ions, and molecular ions. The scheme may also be used for measureing spin imbalances of neutral atoms or atomic ensembles trapped by optical dipole forces. As the proposed method relies on techniques well established in ion trap quantum information processing it is within reach of current technology.}

\begin{document}
\maketitle

\section{Introduction}
Recently developed quantum enhanced precision measurement schemes in atomic physics have the potential to improve classical methods significantly~\cite{Giovannetti:2004}. Entangled states can for instance be used to improve the sensitivity of Ramsey spectroscopy \cite{Bollinger:1996} for frequency estimation or the detection of external fields. Furthermore, sensitivity due to external field noise \cite{Huelga:1997} can be circumvented in turn by carefully designing robust quantum states~\cite{ROOS06}, insensitive to the prevailing correlated noise in e.g. ion trap or optical lattice experiments~\cite{Dorner:2012}, while still improving the sensitivity to the field of interest. Importantly, and in contrast to other emerging quantum technologies such as quantum computing and simulation, improvements already materialise for relatively small quantum systems. This fits especially well to the experimental approach using small trapped crystals of cold ions, where entangled states of up to 14 ions \cite{LEIB05,HAFF2006,Monz2010} have been generated. This control of entanglement makes trapped ions ideal systems for performing precision spectroscopy.


A notable scheme making use of quantum effects is quantum logic spectroscopy~\cite{ROSEN} for realizing atomic clocks with relative frequency uncertainties of a few parts in 10$^{17}$. Here, a "logic" ion such as Be$^+$ is used to read out the quantum state of the Al$^+$ "clock" ion. Slight modifications  and further applications of the scheme have been proposed e.g. for the measurement of the Lamb shift in He$^+$ \cite{HERMA2009}, in the context of optical comb spectroscopy of metal ions \cite{HEMMERLING2011,Schmidt2009} to obtain  spectroscopic data with astrophysical relevance, and for the molecular spectroscopy of N$_2^+$ ions \cite{TONG2010}. In all those cases the information on a successful laser excitation of the clock (or "spectroscopy") ion is conveyed to the logic ion via a collective vibrational degree of freedom.

\begin{figure}
\onefigure{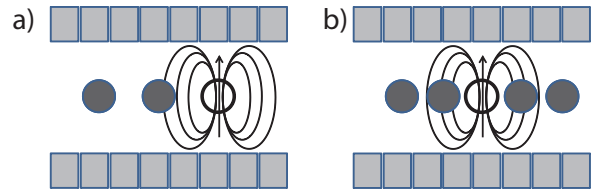}
\caption{Principle of the proposed measurement: a) The Bell state $1/\sqrt{2} \{|\uparrow\rangle_1 |\downarrow \rangle_2 + |\downarrow \rangle_1 |\uparrow\rangle_2\}$ with two $^{40}$Ca$^+$ ions (dark grey circles)  is trapped in a micro trap (light grey) in close vicinity to a third ion X whose spin is to be measured (white circle). The magnetic dipole field affects mostly the neighboring ion and leads to a phase shift which can be detected in a parity signal (see text). b) The Greenberger Horne Zeilinger state $\frac{1}{\sqrt{4}}|\uparrow_1,\downarrow_2,X,\downarrow_3,\uparrow_4\rangle+e^{i\phi}|\downarrow_1,\uparrow_2,X,\uparrow_3,\downarrow_4\rangle$ is used to measure the spin state of ion X in the middle (white circle).}\label{scheme}
\end{figure}

Here, we propose a fundamentally different approach, which does not depend on a collective mode of motion, shared between the measurement ion and the atom or ion to be measured. Instead, direct coupling of the magnetic field, generated by the spins of the atoms or ions to be measured, to the spins of the measurement ions is used. In particular, we show that the field gradient of a source as small as a single electronic spin can be detected over a distance of a few $\mu$m using a specially designed quantum state of two ions that is insensitive to magnetic field noise.

\section{Quantum enhanced sensing of a magnetic field gradient}
Here, we focus on the details of our method, discussing the most simple case of a spin detection in $^{40}$Ca$^+$. After outlining the details of the detection features, we exemplify the spectroscopy of magnetic properties of atomic or molecular ions which do not allow optical excitation or detection. Finally, one of the further applications is sketched, the measurement of neutral atom magnetic moments.

The method is based on well developed methods in ion traps quantum processors \cite{Blatt2008}. For the implementation one would prefer a micro structured segmented ion trap where ions can be moved along the trap axis and positioned with high accuracy and second, the small trap dimensions allow for high confinement with small inter-ion distances \cite{KIELPINSKI2002,Blakestad2009,Huber2010,WALTHER11}. In the first step, a Bell state $\Psi^+$ with two $^{40}$Ca$^+$ ions is generated
\begin{equation}
    \Psi^+ = |\uparrow\rangle_1 |\downarrow \rangle_2 + e^{i\phi}|\downarrow \rangle_1 |\uparrow\rangle_2
\end{equation}
where the phase $\phi$ is chosen equal to zero and where $|\downarrow \rangle$ denotes the S$_{1/2}$ m=-1/2 Zeeman level and $|\uparrow\rangle$ the m=+1/2, respectively and we omitted the normalization factor for clarity. The index number corresponds to the ion 1 and 2. This state may be generated in a Cirac-Zoller type \cite{SCHMIDT2003}, a Moelmer-Soerenson type \cite{BEN08} or a geometric \cite{LEIBFRIED2003a} type of gate operation, depending on the experimental preferences. Gate operations are reaching state fidelities of about 99$\%$. In the specific case of the Moelmer-Soerensen operation, a bichromatic laser field generates a Bell state like $|S_{1/2}\rangle_1 |D_{5/2}\rangle_2 + |D_{5/2}\rangle_1 |S_{1/2}\rangle_2$, where the Bell state is encoded in the $S$ ground and the metastable $D_{5/2}$ states. Important for the proposed scheme is then to transfer the entanglement from the optical qubit transition, a superposition of the S$_{1/2}$ m=-1/2 and D$_{5/2}$ m=+1/2 state into the in Ca$^+$ ground state Zeeman superpositions $|\uparrow\rangle, |\downarrow\rangle$. It had been shown experimentally that this Bell state $\Psi^+$ exhibits a coherence time well above 10~seconds, even though no special care had been taken by e.g. magnetic shielding \cite{HAEFF2005}. The reason for this striking property comes from the advantages of decoherence-free subspaces: the linear Zeeman shift of the two ions leads to phase shifts acting in opposite directions, and as both ion are at a distance of a few $\mu$m experiencing very similar magnetic field noise, this leads to an almost perfect cancelation of differential phase shifts. An other possible source of decoherence, spontaneous decay, is fully excluded since sublevels of the ground state are used. Once the entangled state of both "logic" ions is generated one approaches in the trap - by addressing the control voltages -  the third ion X thats magnetic field is to be determined. Trap frequencies $\omega_{z}/(2\pi)$ of 5 to 10~MHz are typically reached and allow compressing the three ion crystal to reach inter ion distances of a few~$\mu$m. The magnetic dipole field scales with d$^{-3}$, resulting in a gradient over the Bell state that leads to differential Zeeman shifts which results in a rotation of the phase angle $\phi$, see Fig.~1a. After an interaction time $t$, one applies a $(\pi/2)$-pulse on the logic ions, transfers the entanglement from the $\{|\uparrow\rangle,|\downarrow\rangle\}$ states to the $\{S_{1/2}, D_{5/2}\}$ bases and reads out the laser induced fluorescence when applying resonant light on the dipole transitions from the $S_{1/2}$ to the $P_{1/2}$ state near 397~nm, and $D_{3/2}$ to the $P_{1/2}$ state near 866~nm, respectively. The fluorescence detection allows the determination of the quantum state of the two Ca$^+$ ions. Repeating the above sequence many times allows to determine the average values $p_{\uparrow}$ and $p_{\downarrow}$ for both ions from where the parity $P$ of the state $P=(p_{\uparrow \uparrow} + p_{\downarrow \downarrow}) - (p_{\uparrow \downarrow} + p_{\uparrow \downarrow})$ is calculated. The shift of the phase $\phi$ of the Bell state leads to sinusoidal oscillations from P($\phi=0$)=+1 to P($\phi=2\pi$)=-1.

\begin{figure}
\onefigure{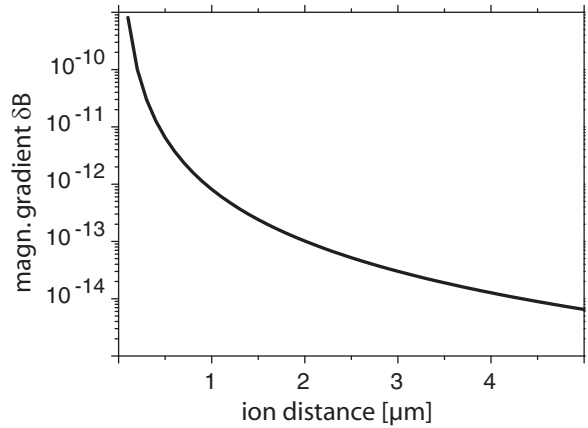}
\caption{Magnetic gradient between ion 1 and ion 2 as a function of their distance when ion 3 is spin polarized in z-direction.}\label{grad}
\end{figure}

Next, we estimate the relevant parameters such as measurement time, noise sources and detection limits, for supporting the feasibility of the scheme under state-to-the-art experimental conditions. The three-ion crystal distance in a linear configuration reads as $d_{12} = 1.077 (e^2/(4\pi \epsilon_0 m \omega_z^2)^{1/3}$ where $m$ denotes the mass of the ions which are confined in axial direction in the trap with frequency $\omega_z$ \cite{JAMES}. A distance of $d_{12}$=1.03~$\mu$m is reached for $\omega_z/2\pi$=10~MHz which increases to 1.63~$\mu$m for $\omega_z/2\pi$=5~MHz. A magnetic dipole field of an atom or ion is
\begin{equation}
    \overrightarrow{B}(\overrightarrow{r})=\frac{\mu_0}{4\pi} \frac{3\overrightarrow{r}(\overrightarrow{m} \cdot \overrightarrow{r}) - \overrightarrow{m} r^2} {r^5}
\end{equation}
where $\mu_0$ denotes the permeability constant and $\overrightarrow{m}$ the magnetic dipole moment \cite{Jackson}. Let us assume further that the quantization axis is determined by an external homogenous field B$_z$ pointing in z-direction along the trap axis, the magnetic field $B_z(x,y,z)$ results in
\begin{equation}
    |B_z|(x,y,z)= \frac{3 \mu \mu_0}{4 \pi}
    \frac{(3z^2 - 1) (x^2 + y^2 +
     z^2)}{(x^2 + y^2 + z^2)^{5/2}}
\end{equation}
with $\mu_0 = 4\pi \cdot 10^{-7} H/m$ and the intrinsic magnetic moment $\mu$ = -9284.764 $\cdot 10^{-26}$ J/T in SI-units. From this equation one finds that the field generated by the third ion X drops from $7.8 \cdot 10^{-13}$~T at the location of the second ion (at 1.03~$\mu$m distance) by almost one order of magnitude to $9.7 \cdot 10^{-14}$~T at the location of the first one (at 2.06~$\mu$m distance) since the magnitude $|B(r)|$ is scaling with $r^{-3}$. For sensing the gradient field of ion X, see Fig.~2, we apply an external gradient of opposite direction and same magnitude such that for the state $|\uparrow\rangle_X$ both contributions cancel and the entangled state parity remains constant with time. For the state $|\downarrow\rangle_X$, however, both contributions add up to $\delta B = 2(B(z_{ion2})-B(z_{ion1}))$ and the parity signal exhibits a phase evolution. For the above example a $\delta B$ = 6.8 $\cdot 10^{-13}$~T would lead to a sinusoidal rotation in $P(\phi)$ from +1 to -1 in 26~s. One may detect smaller variations of the  parity signal close to its zero-crossing of $\pm 30\%$ with an evolution time of 5~s when the spin of the ion X is flipped from $|\uparrow\rangle_X$ to $|\downarrow\rangle_X$. Assuming quantum projection noise as dominating source one would reach a signal-to-noise ratio of 2 after about only 10 repetitions of the experimental sequence, and with a total measurement time of less than 60~s. For first tests of the method, one might use a three-ion crystal of $^{40}$Ca$^+$ ions, where the spin in the third ion can be aligned with optical pumping. In such experiments, one might improve the long-term stability of the apparatus. Reducing the ambient magnetic gradient fluctuations below $10^{-13}T/\mu m$ for the relevant time scale may require passive magnetic shielding and active control. The technique using entangled, designed states of a two-ion crystal are an alternative promising way to overcome the ambient magnetic field noise, as compared to lock-in methods which are aiming for the similar sensitivities \cite{KOT11}.

\section{Proposed applications}
The new method might be applied for yielding spectroscopic data of the magnetic sublevels in the case of ions without accessible dipole transitions for optical cooling and detection. The substructure of molecular ions could be investigated with the detection technique as well: trapped in a three-ion crystal together with the entangled Ca$^+$ pair, an excitation of the molecular ion is tested with a laser pulse of defined frequency and duration. When the excitation of molecular transition was successful, the molecular state was changed and the corresponding modification of the magnetic moment is detected from the parity signal of the entangled $^{40}$Ca$^+$ ions. The proposed method extends the idea of 'designer ions' \cite{ROOS06} which has been used to investigate accurately frequency shifts on an optical transition frequency - to ion crystals with multiple species ions. The method may be compared to destructive measurement approaches realized with large mixed ion crystals of $^{X}$N$_2^+$ ions sympathetically cooled in a $^{40}$Ca$^+$ crystal. A successful molecular transition event was either detected by the subsequent ion loss \cite{TONG2010} or by charge transfer processes \cite{STA2010}. As compared with a recently proposed method, we do not require strong magnetic gradient fields or optical forces for the detection \cite{MUR11}.

The method is not limited to detect magnetic properties of atomic or molecular ions confined in a Paul trap, as the interaction does not require a coupling to a common vibrational mode. Thus, the Ca$^+$ two-ion crystal might be positioned as a local probe in a ensemble of neutral atoms and sense the magnetic properties, e.g. for detecting quantum magnetic phases. The advantage of this system is that ions might be approached much closer to the neutral atom spins, which increases the sensitivity. In this way, the two-ion entangled state would directly serve to detect the \emph{difference} of magnetic field at its both sites, while rejecting any common magnetic field fluctuations. In recent experiments the operation of combined traps for ions and neutral atoms has been demonstrated \cite{ZIP10,SCHM10} and it appears as logical next step to use the advantages of entangled ion crystals also here. Especially when the neutral atoms are trapped in a double-well potential, the entangled ion pair in the middle would sense the imbalance of magnetic field generated by the right and the left hand side, which are in in typical realizations a few $\mu$m apart. To exemplify the feasibility of the scheme we assume a double-well separated by 4.4$\mu$m, like the experimental realization with optical dipole potentials and crossed beams under 9$�$, Ref.~\cite{GATI06}. The two-ion crystal with an inter-ion distance of 3.5$\mu$m would perfectly fit in between the degenerate gas in the potential well, and the ions are not immersed in the gas, such that collisions and heating effects are largely avoided. When we assume an interaction time of 2.5~s, and a magnetic gradient over both ions caused by an imbalance of only singe atomic dipole contribution would be about $\delta B$ = 13 $\cdot 10^{-12}$~T would turn out in a $\pm 30\%$ parity modulation signal. The signal directly reveals the difference of atom population $\Delta N$ in the two wells. The sensitivity reaches the single atom level $\Delta N$=1 and could be used to detect number fluctuations at the quantum noise limit \cite{GROSS10} and the Josephson dynamics of coupled potentials \cite{GER2011}.

Let us compare our scheme with conventional magnetic field sensing techniques. Conventional magnetic sensing operates with SQUID, giant magnetoresistance or fluxgate sensors. The typical size of such devices ranges from a few mm to about 100$\mu$m and pairs or even arrays of SQUID detectors are used for a measurement of magnetic field gradients. However, such devices do not reach the single spin level. Magnetic resonance force microscopy, on the other hand, has proven to reach single electron spin sensitivity at a measurement distance of up to 100~nm \cite{RUGAR2004}. Alternatively, nitrogen color centers (NV) in diamond serve as magnetic field sensors \cite{MAZE2008}. The long coherence time of the single NV quantum state is critical for reaching the best sensitivity \cite{MAZE2008,WALDHERR2012}. The width of the resonance between Zeeman shifted levels leads to a  sensitivity limit of about 80 10$^{-6}$ T/nm for spin imaging \cite{BALAS2008}. A sensitivity, sufficiently high to detect a single spin, is reached in such device at a distance of about 5~nm \cite{GROTZ2011,BALAS2008}.  In contrast to these methods, the proposed scheme uses i) entangled ions in a quantum state where only gradients of the magnetic field lead to a signal and where ii) the long coherence time of ions allows an improved sensitivity.

\section{Outlook}

One might further extend the method to linear ion crystals with N-particle entanglement. One would embed an ion X into the linear Ca$^+$ crystal. The phase evolution of the N-particle entangled state, as measured by the parity signal would be strongly influenced by the state of spin X and the corresponding local field gradients due to its magnetic moment. While there are certainly several experimental problems due to the complexity of this scheme, two advantages are clearly visible: the ion-distances become even smaller when the number of ions increases, which leads to stronger magnetic field gradients, and the multi-particle entanglement is leading to faster parity rotations \cite{LEIB05} allowing a more accurate detection. For this case the N-ion Greenberger Horne Zeilinger (GHZ) state can be used, which can also be prepared in a decoherence-free, and thus long lived version i.e.
\begin{equation}
    \Psi^+ = \frac{1}{\sqrt{4}} \hspace{2mm} |\uparrow_1, \downarrow_2, X, \downarrow_3, \uparrow_4 \rangle + e^{i\phi}|\downarrow_1, \uparrow_2, X, \uparrow_3, \downarrow_4 \rangle
\end{equation}
Note, that the generation of the GHZ state does not require individual addressing of  $^{40}$Ca$^+$ ions with a laser beam. The electronic transitions of the ion in the central position of species X are not resonant to that light used for the $^{40}$Ca$^+$ qubits. Finally, any spin X would be detected in a doubled phase shift induced by constructive addition of the opposite Zeeman effects to the ions 2 and 3, see see Fig.~1b.

In conclusion, the proposed method extends the toolbox of quantum metrology. The entangled two-ion crystals serve as a very sensitive local magnetic field probe that  directly reveals the difference of fields at the ion locations while any common magnetic noise contributions are suppressed. This quantum sensor may find applications for spectroscopy of ions of other species with transitions that are not easily accessible, for molecular ion spectroscopy and to detect the magnetic fields of neutral atoms when they act as sources of magnetic dipole fields.

\acknowledgments
This work has been supported by the European Commission (IP-AQUTE).

\bibliographystyle{prsty}
\bibliography{MagGrad_bib}
\end{document}